\documentclass{article}
\usepackage{spconf,amsmath,graphicx}

\usepackage{csquotes}
\usepackage{color, colortbl}
\usepackage{soul}

\usepackage{amsmath,amsfonts,bm}

\def\eqref#1{equation~\ref{#1}}

\def\1{\bm{1}}

\DeclareMathAlphabet{\mathsfit}{\encodingdefault}{\sfdefault}{m}{sl}
\SetMathAlphabet{\mathsfit}{bold}{\encodingdefault}{\sfdefault}{bx}{n}
\newcommand{\tens}[1]{\bm{\mathsfit{#1}}}

\def\tF{{\tens{F}}}

\def\tI{{\tens{I}}}

\def\tW{{\tens{W}}}

\title{pan-sharpening with color-aware perceptual loss and guided re-colorization}
\name{Juan Luis Gonzalez Bello*, Soomin Seo*, and Munchurl Kim\thanks{* Both authors contributed equally to this work.}}
\address{Korea Advanced Institute of Science and Technology (KAIST)}
\begin{document}
\ninept
\maketitle
\begin{abstract}
In remote sensing, \enquote{pan-sharpening} is the task of enhancing the spatial resolution of a multi-spectral (MS) image by exploiting the high-frequency information in a panchromatic (PAN) reference image.
We present a novel color-aware perceptual (CAP) loss for learning the task of pan-sharpening. Our CAP loss is designed to focus on the deep features of a pre-trained VGG network that are more sensitive to spatial details and ignore color information to allow the network to extract the structural information from the PAN image while keeping the color from the lower resolution MS image. Additionally, we propose \enquote{guided re-colorization}, which generates a pan-sharpened image with real colors from the MS input by \enquote{picking} the closest MS pixel color for each pan-sharpened pixel, as a human operator would do in manual colorization. Such a re-colorized (RC) image is completely aligned with the pan-sharpened (PS) network output and can be used as a self-supervision signal during training, or to enhance the colors in the PS image during test. We present several experiments where our network trained with our CAP loss generates naturally looking pan-sharpened images with fewer artifacts and outperforms the state-of-the-arts on the WorldView3 dataset in terms of ERGAS, SCC, and QNR metrics.
\end{abstract}


\section{Introduction}
\label{sec:intro}
Satellite imagery are photographs of the Earth or other planets obtained by imaging satellites, containing not only the RGB data which is visible for the human vision but also near-infrared (NIR) and shortwave infrared (SWIR) data. Satellite imagery is broadly utilized in geology, agriculture, environmental monitoring, and many other applications. Imaging satellites often record multiple spectral bands with different spatial resolutions due to physical constraints such as sensor resolution and bandwidth limitations. In general, the image data from satellites include a high-resolution (HR) panchromatic (PAN) image and a low-resolution (LR) multi-spectral (MS) image for the same scene. Pan-sharpening or pan-colorization is the task of generating a pan-sharpened (PS) multi-spectral image which has the same spatial resolution as the PAN image, by fusing the high-frequency details from the latter and the color information from the MS image.

With recent advances in deep learning for image processing, several works on pan-sharpening have been proposed that incorporate learning models with deep convolutional neural networks (DCNN) \cite{pannet, dsen2, s3, vdsrpan, targetpan, pnn, srcnnpan, progressivepan, fullypan, densepan, unetpan}. However, most of the deep learning-based pan-sharpening methods simply regularize the network by minimizing a spectral loss between the generated output and a pseudo-ground-truth MS image, without considering the inherent image distribution difference between the PAN and MS images. PAN images generally cover a wide range of wavelengths by merging a broad spectrum of visible lights into a single channel image. Therefore, even if the MS bands are converted into a single-channel gray-scale image, its luminance will considerably differ from the PAN image. For examples, certain objects that appear bright in an MS image (water, etc.) can appear dark on a corresponding PAN image or vice-versa (trees, grass, etc.). This inherent luminance difference between PAN and MS images generates not only mismatching luminance values, but also opposite directions of gradients between them, which hinders a DCNN from properly learning the task of pan-sharpening.

To resolve the aforementioned issue, we propose a novel cost function called color-aware perceptual (CAP) loss, which is specifically designed to focus on the features of a pre-trained VGG network that are more sensitive to high-frequency details and less sensitive to color differences. The proposed CAP loss allows the network to preserve details even if the gradient directions do not match between PAN and MS images, thus resulting in enhanced details and less artifacts around the edges. In addition, we propose a \enquote{guided re-colorization} module which allows the network outputs to have even more similar colors with the input MS images. The module generates a re-colorized PS image with real colors from the input MS image by picking the closest MS pixel color in the neighboring area. This re-colorized image is further combined with the high-frequency contents of the initial pan-sharpened output to generate the final re-colorized output PS image. Guided re-colorization can be used as a post-processing step at testing time, or as a self-supervision signal during training. The presented experiments show that our method achieves the state-of-the-art performance compared to the previous works in terms of structural information preservation. Furthermore, we present a qualitative comparison from which our network is shown to effectively preserve edge details with less artifacts.

\begin{figure*}
  \centering 
  \includegraphics[width=0.99\textwidth]{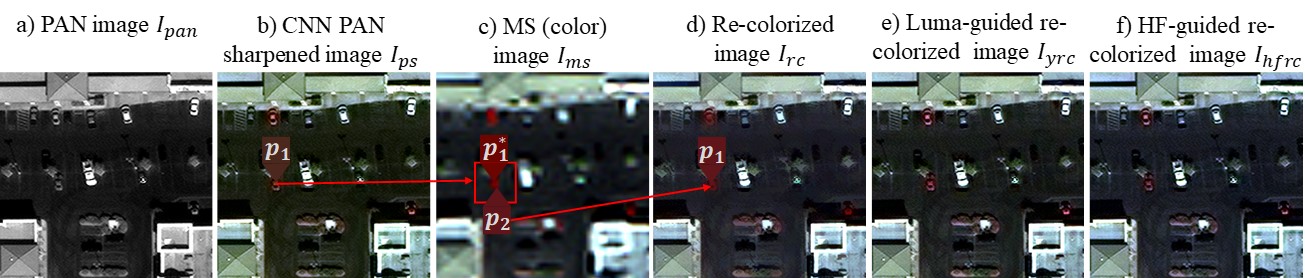}
  \vspace*{-3mm}
  \caption{Guided re-colorization. A \enquote{real color} is searched in a local MS-color image window for each pixel in the CNN-colorized image. For example, to re-colorize the pixel at location $p_1$ in (b), the closest color is searched in a local window in (c) where the pixel neighbor at location $p_2$ has the closest color to $p_1$. The pixel $p_2$ is used to re-color the pixel at location $p_1$ in (d). Even when the selected color is not the optimum (as $p^*_1$), the re-colorized output generates a closer color prediction to the MS image. Lost edge information can be recovered by luma guidance (e) or high-frequency guidance (f).}
  \vspace*{-3mm}
  \label{fig:rc_explain}
\end{figure*}

\section{Related works}
\label{sec:related}

\subsection{Deep learning-based pan-sharpening methods}
Following the success of deep learning-based methods in the image processing field, most recent works have incorporated DCNN architectures that have proven to be effective on the super-resolution (SR) task. Masi \textit{et al.} proposed PNN \cite{pnn}, which is known to be the first work to utilize a DCNN for pan-sharpening, and borrowed its three-layered architecture from the first DCNN-based SR work, SRCNN \cite{srcnn}. The later work of Yang \textit{et al.}, PanNet \cite{pannet}, incorporated a ResNet \cite{resnet} architecture used for classification as their backbone network where a residual connection allows the network to focus on preserving the high-frequency details. PanNet tries to focus even more on high-frequency components by feeding only the high-pass-filtered PAN and MS images as inputs to the network. Similarly, Lanaras \textit{et al.} adopted a deeper SR network, EDSR \cite{edsr}, as the backbone architecture and proposed DSen2 \cite{dsen2} (and a deeper version, VDSen2) for pan-sharpening. Zhang \textit{et al.}, in their BDPN \cite{bdpn}, proposed a bidirectional pyramid network that processes the MS and PAN images in two separate branches, which allows the spatial details from the PAN image to be injected into the spectral information from the MS image to generate the final PS image. Recently, Choi \textit{et al.} proposed an S3 \cite{s3} loss, which can be incorporated into the training of any CNN-based pan-sharpening method to reduce the artifacts by considering the correlation between the PAN and MS images.

\subsection{Perceptual loss}
The perceptual loss, first introduced by Johnson \textit{et al.} \cite{perceptualloss} has been widely adopted as it demonstrated to be a useful tool for training CNNs for several applications, ranging from super-resolution to novel view synthesis and pixel to pixel image transformations. The recent work of Tariq \textit{et al.} \cite{tt} analyzed the correlation between the VGG \cite{vgg} network activations (utilized in perceptual loss) and the human visual system to re-weight the perceptual loss and improve its response for the super-resolution task. In a similar way, we propose our novel color-aware perceptual (CAP) loss, which focuses on those VGG channels that are less sensitive to color (and gradient directions) and more sensitive to structural information. Our CAP loss allows the network to better extract the high-frequency and structural information from the PAN while keeping the colors from the MS.

\section{Method}
\label{sec:method}
In the following subsections, we explain in depth our novel color-aware perceptual loss, designed to focus on the color invariant features of a pre-trained VGG network, and the extensive use of our \enquote{guided re-colorization}, as a new module in our network architecture, a loss term, and as a post-processing step.

\begin{figure*}
  \centering 
  \includegraphics[width=0.95\textwidth]{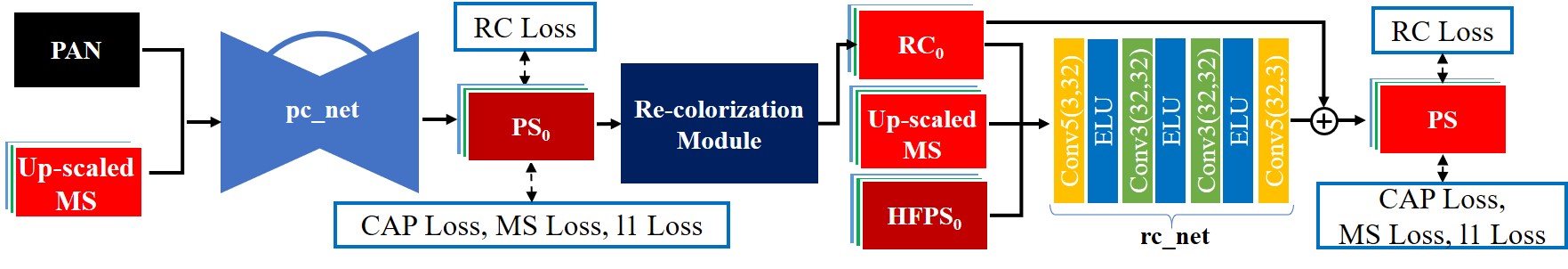}
  \vspace*{-4mm}
  \caption{Network architecture of our proposed pc-net-rc.}
  \vspace*{-3mm}
  \label{fig:network}
\end{figure*}

\subsection{Color-aware perceptual loss}
Given a target or ground-truth image $\tI$ and a predicted image $\tI'$, the vanilla perceptual loss is given by
\begin{equation} \label{eq:p_loss}
l_p = \sum_{l=1}^{L} ||\phi^l(\tI)-\phi^l(\tI')||^2_2
\end{equation}
where $\phi^l$ is the output of the convolutional layers 1, 2 and 3 for $L=3$ of the VGG19, pre-trained for image classification task on ImageNet. Perceptual loss has shown to be more effective for image restoration than the plain $l1$ or $l2$ losses. The reason behind its effectiveness is that instead of penalizing a single intensity value at a certain pixel location, perceptual loss penalizes the relationship between the target pixel and the neighboring pixels given by the layer's receptive field and learned filter parameters. Such receptive field size grows as we go to deeper layers.

Since the early work of the Alexnet \cite{alexnet}, it has been observed that certain shallow convolutional filters are more sensible to colors than others. Inspired by this observation, our proposed color-aware perceptual (CAP) loss is designed to assign less contribution or less weight to those feature channels that are more sensible to color for each of the $L$ VGG19 layers when computing the perceptual loss between the PAN image and the pan-sharpened (PS) image. We propose to obtain such weights by measuring the feature differences between the corresponding color MS ($\tI_{ms}$) image and a grayscale-inverted MS image ($\tI^{-1}_{gms}$). The greater the difference, the more sensible the feature is to color and brightness (which is the reason behind inverting the grayscale version of the MS image). Color inversion also makes our CAP loss sensitive to gradient inversions, a common feature between MS and PAN images. The mean feature difference is then fed into an exponential function with an adjustable parameter $\gamma$. This process is described by
\begin{equation} \label{eq:gray_inverted}
\tI^{-1}_{gms} = 1 - (\tI^R_{ms} + \tI^G_{ms} + \tI^B_{ms}) / 3
\end{equation}
where $\tI^R_{ms}$, $\tI^G_{ms}$, $\tI^B_{ms}$ are the RGB components of the MS image normalized to a range from 0 to 1. The color-aware perceptual weights for each channel of each layer $l$, $\tW^l_{cap}$, are then given by
\begin{equation} \label{eq:w_cap}
\tW^l_{cap} = e^{-\gamma(1/N)\sum_{n}^{N} |\phi^l(\tI^{-1}_{gms})-\phi^l(\tI_{ms})|}
\end{equation}
where $\gamma$ was empirically set to 4 to neglect most features that are sensible to color. Then, given a PAN image $\tI_{pan}$ and a CNN pan-sharpened image $\tI_{ps}$, our color-aware perceptual loss is given by
\begin{equation} \label{eq:cap_loss}
l_{cap} = \sum_{l=1}^{L}||\tW^l_{cap}\odot(\phi^l_{m_l}(\tI_{pan})-\phi^l_{m_l}(\tI_{ps}))||_1
\end{equation}
where $m_l = [7, 5, 3]$ indicates a max-pooling size applied for the $l$ layer feature that was adopted to provide means of shift invariance to our CAP loss, needed to better handle the MS-PAN miss-alignments. 

While the CAP loss enforces high frequency details on $\tI_{ps}$, additional loss terms are needed to enforce color fidelity. For this, we use the plain perceptual loss and $l_1$ loss. Then, our fidelity loss function is given by
\begin{equation} \label{eq:total_loss}
l_{f} = \alpha_{cap}l_{cap}(\tI_{pan}, \tI_{ps}) + \alpha_{ms}l_p(\tI_{ms}, \tI_{ps\downarrow}) + \alpha_{l1}l_1(\tI_{ms}, \tI_{ps\downarrow})
\end{equation}
where $\tI_{ps\downarrow}$ indicates a downscaled PS image to the MS resolution. The $l_1$ loss is the mean absolute error or $1/N\sum_{n}^{N}|\tI_{ms}-\tI_{ps\downarrow}|$. The constants $\alpha_{cap}$, $\alpha_{ms}$ and $\alpha_{l1}$ were empirically set to 0.9, 0.01 and 1.0 respectively. Note that more weight is assigned to the CAP loss and this is possible thanks to its color invariant features.

\subsection{Guided re-colorization}
Once a CNN has been trained to colorize a pan image, a post-processing step can further be incorporated by reflecting how a human user would correct CNN-generated PS images. Our guided re-colorization (RC) works on the assumption that a CNN-colorized image can be further improved in terms of spectral fidelity or color by changing the CNN-generated color with the closest MS color in a local window. This may be done in a similar way as a human operator would pick the closest color in the MS-color image given a sub-optimally colorized PAN image. As shown in Fig. \ref{fig:rc_explain}-(b,c,d), for each pixel at location $p_1=(x,y)$ in the colorized PAN image, a \enquote{real color} is searched in a local window in the bilinearly up-scaled MS image $\tI_{ms\uparrow}$. The closest color is selected based on the euclidean distance in the RGB color space. We denote this re-colorization operation by
\begin{equation} \label{eq:rc_eq}
\tI_{rc}=N(\tI_{ps}, \tI_{ms\uparrow}, w)
\end{equation}
where $w$ is the window size. However, such operation can degrade the high frequency details as shown in Fig. \ref{fig:rc_explain}-(d). To revert this effect, we adopted two different approaches: (1) The raw re-colorized image $\tI_{rc}$ can be transformed to the YCbCr color space and its luma or Y-channel replaced with the Y-channel of the CNN-colorized image $\tI_{ps}$ to generate a luma-guided re-colorized image $\tI_{yrc}$ as shown in Fig. \ref{fig:rc_explain}-(e); (2) A high-frequency guided re-colorized image $I_{hfrc}$ can be generated by subtracting the low-frequency (LF) information from the CNN-colorized image, and adding the LF information of $\tI_{rc}$ image as described by 
\begin{equation} \label{eq:hfrc_eq}
\tI_{hfrc}=\tI_{ps} - \tI_{ps}*\tF_{avg} + \tI_{rc}*\tF_{avg}
\end{equation}
where $\tF_{avg}$ is the averaging filter (with a size set to 5x5 throughout this paper). An example of this operation is shown in  Fig. \ref{fig:rc_explain}-(f). These operations (1 and 2) can recover the high-frequency details thanks to the fact that the pan-sharpened image $\tI_{ps}$ and the re-colorized image $\tI_{rc}$  are perfectly aligned, which is not the case for the MS image. The re-colorized module is implemented in our network architecture as an intermediate block, as explained in the next section. While the high-frequency guided image $\tI_{hfrc}$ is used as a post-processing step, the luma-guided RC image $\tI_{yrc}$ is used as a self-supervised signal to enforce the generation of real-colors at target resolution (instead of CNN hallucinated color) by adopting the RC loss term $l_{rc}$ as
\begin{equation} \label{eq:rc_loss}
l_{rc}=||\tI_{ps} - \tI_{yrc}||_1
\end{equation}
The use of $\tI_{yrc}$ in the RC loss showed to be more stable for training than $\tI_{hfrc}$, as the first keeps all the edge information from the luma channel. In contrast, in the $\tI_{hfrc}$, the edges can slightly change depending on the average filter $\tF_{avg}$ size. Finally, the total loss is the sum of the fidelity loss $l_{f}$ and the RC loss $l_{rc}$.

\subsection{Network architecture}
Our network architecture, which we call the pc-net-rc, is depicted in Fig. \ref{fig:network}. In its first stage, an initial pan-sharpened image $\tI^0_{ps}$ is estimated by the pc\_net, which is an auto-encoder network with skip connections (similar to an U-Net \cite{unet}). This type of architecture allows the network to account for global and local information, which is not only useful for colorizing individual pixels, but also for handling the PAN-MS misalignment. In the second stage, the rc\_net takes as input the re-colorized version of $\tI^0_{ps}$, which is $\tI^0_{rc}=N(\tI^0_{ps}, \tI_{ms\uparrow}, w\text{=}3)$, the original MS image and the high-frequency component of $\tI^0_{ps}$. The rc\_net is composed of four conv-elu blocks and a residual connection with $\tI^0_{rc}$. The final pan-sharpened image $\tI_{ps}$ is then described by
\begin{equation} \label{eq:pc_net}
\tI^0_{ps}=pc\_net(\tI_{pan}, \tI_{ms\uparrow}) 
\end{equation}
\begin{equation} \label{eq:rc_net}
\tI_{ps}=\tI^0_{rc} + rc\_net(\tI^0_{rc}, \tI_{ms\uparrow}, \tI^0_{ps} - \tI^0_{ps}*\tF_{avg}) 
\end{equation}

\begin{figure*}
  \centering 
  \includegraphics[width=0.95\textwidth]{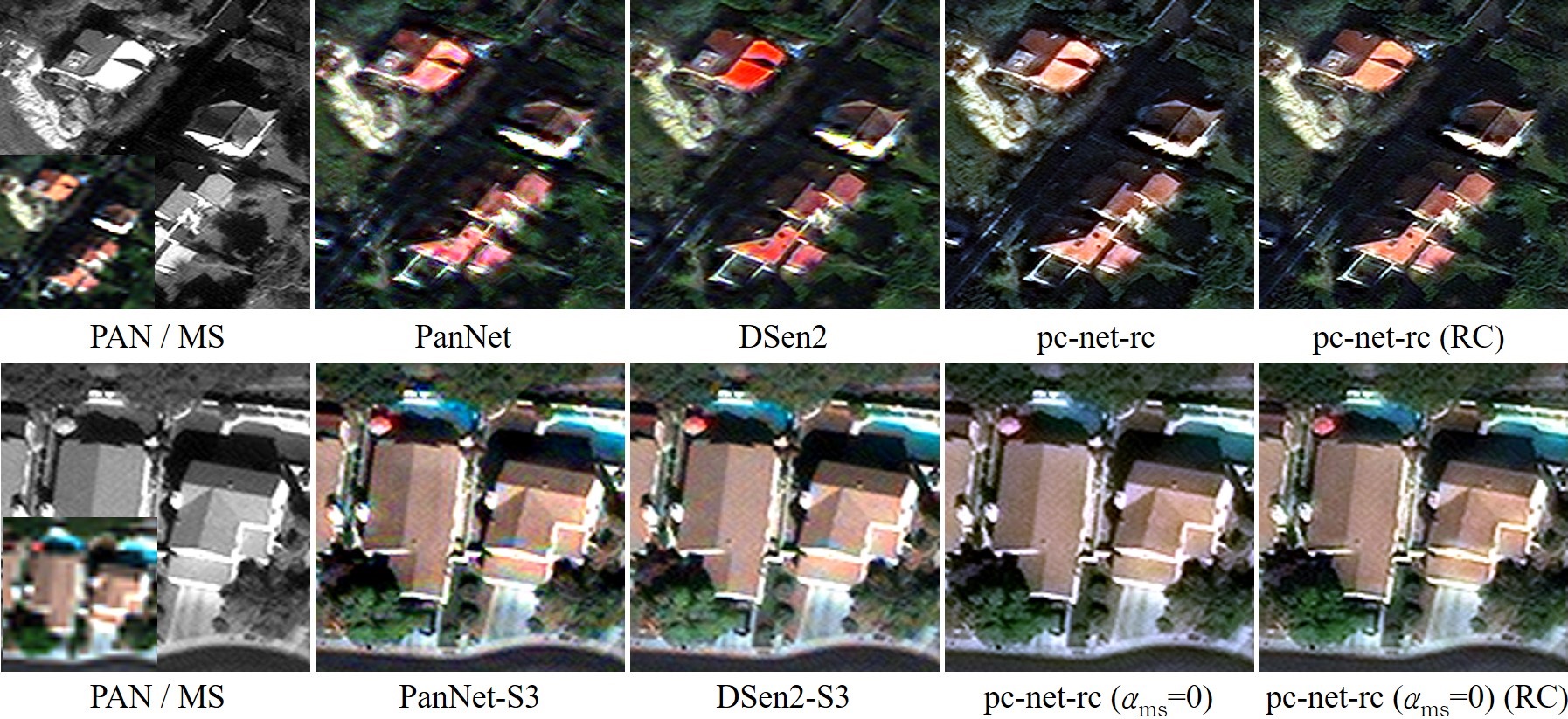}
  \vspace*{-4mm}
  \caption{Qualitative comparison among existing methods. Top row: methods with low ERGAS. Bottom row: methods with high SCC.}
  \vspace*{-4mm}
  \label{fig:results_comp}
\end{figure*}

\begin{figure}
  \centering 
  \includegraphics[width=0.48\textwidth]{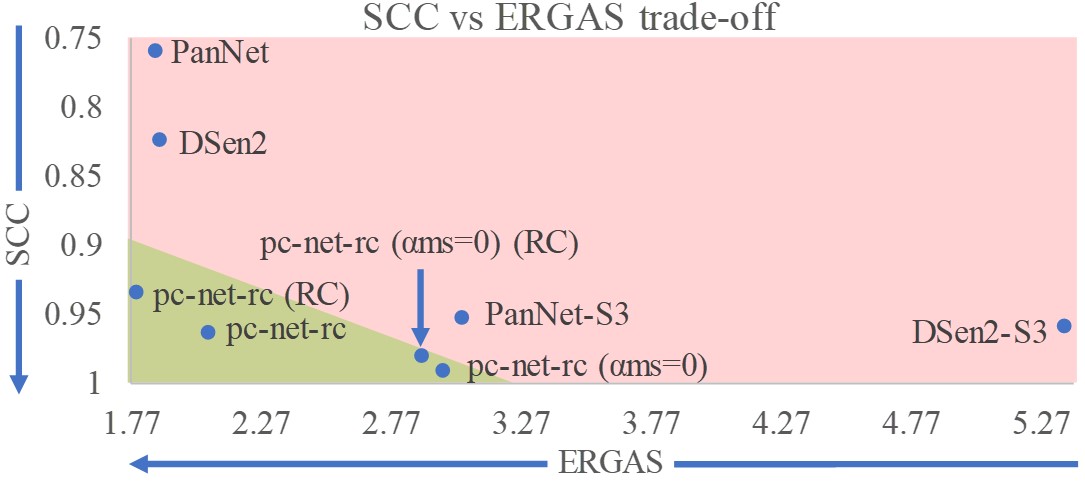}
  \vspace*{-8mm}
  \caption{Our proposed network with CAP loss (in green area) presents a better SCC-ERGAS trade-off than previous works (in red area).}
  \vspace*{-5.5mm}
  \label{fig:trade_off}
\end{figure}

\section{Experiments and results}
\label{sec:experiments}
We performed extensive experiments and ablation studies on the WorldView-3 dataset that support the effectiveness of our method. The WorldView-3 dataset provides 0.3m GSD (ground sample distance) PAN and 1.2m GSD MS images. It also provides 0.3m GSD PS images but they are only used for comparison, as they are not good enough to be used as ground-truth training data. Only the RGB channels are used for all our experiments. The WorldView-3 dataset used for training and testing was obtained from the SpaceNet Challenge \cite{spacenet}.

\begin{table}[t]
    \small
    \centering
    \begin{tabular}{lcccccccc}
\hline
Method                 & ERGAS$\downarrow$ & SCC$\uparrow$  & QNR$\uparrow$ \\  
\hline
MS-Bilinear            & 1.831 & 0.290  & 0.761 \\
PAN                    & 7.687 & 1.000  & 0.777 \\
\hline
Provided PS            & 3.718 & 0.974 & 0.896 \\ 
PanNet \cite{pannet}   & \underline{1.894} & 0.794 & 0.878 \\
PanNet-S3 \cite{s3}    & 3.044 & 0.952 & 0.923 \\
Dsen2  \cite{dsen2}    & 2.026 & 0.855 & 0.893 \\
Dsen2-S3 \cite{s3}     & 5.354 & 0.959 & 0.915 \\
\hline
pc-net-rc              & 2.066 & 0.963 & \textbf{0.931} \\
pc-net-rc (RC)         & \textbf{1.786} & 0.934 & 0.919 \\
pc-net-rc ($a_{ms}=0$)  & 2.970 & \textbf{0.991} & 0.925 \\
pc-net-rc ($a_{ms}=0$)(RC)  & 2.889 & \underline{0.980} & \underline{0.927} \\
\hline
pc-net* ($a_{ms}=0$) & 3.180 & 0.990 & 0.914 \\
pc-net-rc* ($\gamma=0$) & 4.267 & 0.996 & 0.801\\ 
pc-net-rc* ($a_{cap}=0$) & 1.388 & 0.253 & 0.777 \\
\hline
\end{tabular}

    \vspace*{-2mm}
    \caption{Quantitative comparison. Our pc-net-rc outperforms the SOTAs by a considerable margin in all metrics. $\uparrow\downarrow$ indicate the better performance. \textbf{Best} and \underline{second best} metrics are highlighted, * are not considered as present an undesirable SCC-ERGAS trade-off.}
    \vspace*{-4.5mm}
    \label{tab:table_comp}
\end{table}

\subsection{Implementation details}
We trained our network for 50 epochs with a mini-batch size of 8 by the Adam \cite{adam} optimizer with its default settings. The initial learning rate was set to $5\text{x}10^{-5}$ and halved at epochs 30 and 40. The WorldView-3 dataset was pre-processed to discard images where more than 5\% of the pixels are 0's, yielding a total of 11,804 PAN-MS image pairs. We adopted random data augmentations on-the-fly for all networks, such as horizontal and vertical flips, cropping, and changing luminosity and color brightness. The last two augmentation operations allowed for better performance in terms of ERGAS for all networks. For training the DSen2 (S3) and PanNet (S3), the MS-PAN image pairs were created by the methods suggested in their works \cite{pannet, dsen2, s3}, in which PAN and MS images are down-sampled and the original MS images are used as pseudo-ground-truth. In contrast, we trained our method with the original MS and PAN images.

\subsection{Results on WorldView-3}
We quantitatively compare our proposed method in Table \ref{tab:table_comp} with bilinear upsampled MS images, WorldView-3 provided PS images, and other deep-learning based methods as PanNet \cite{pannet}, DSen2 \cite{dsen2} and their S3 \cite{s3} variants. We use three popular metrics to evaluate the performance of pan-sharpening methods: ERGAS \cite{ergas}, for measuring spectral distortion; SCC \cite{scc}, for measuring spatial distortion; and quality not requiring a reference (QNR) \cite{QNR} metric that considers both spatial and spectral information. We used guided re-colorization as a post-processing step to further improve the spectral quality of our results, denoted by (RC), and achieved the lowest ERGAS. It is also seen in Table \ref{tab:table_comp} that our method trained with $a_{ms}\text{=}0$ achieved the highest SCC score. Additionally we show the results for our method trained with $a_{cap}\text{=}0$ (no CAP loss between PAN and PS images) and $\gamma=0$ (that is, the CAP loss becomes the plain perceptual loss between PAN and PS images), which proves the importance of our CAP loss for exploiting the high frequency details in the PAN images. We also show the results for the pc-net only, which achieves a higher ERGAS versus its pc-net-rc counterpart, proving the usefullness of the rc-net to improve spectral quality. As depicted in Fig. \ref{fig:trade_off}, our method shows a higher SCC than previous methods while achieving a lower ERGAS. 
The effect of such trade-off is observed in Fig. \ref{fig:results_comp}, where the results from PanNet and DSen2 show limited performance on the regions with high MS-PAN mis-alignments, and exhibit blurred edges compared to the PAN input. Their S3 variants show higher details, but some of their edges are not clear and yield color artifacts in homogeneous areas. In contrast, our results show strong edges without any color artifacts. This confirms that our proposed CAP loss is effective for maintaining high-frequency details from the reference PAN input.

\section{Conclusion}
\label{sec:conclusion}
We proposed a novel color-aware perceptual loss which is effective for focusing on details. Our proposed loss can be adopted into any type of pan-sharpening network. Furthermore, we have proposed a guided re-colorization module as a post-processing step which helps the generated output to have closer colors to those of the input MS image. Our guided re-colorization module can be easily adopted as well to any other works to alleviate the spectral gap with input MS images. Through extensive experiments, we have shown that our network produces better pan-sharpened outputs than previous methods in terms of both spectral and spatial based metrics, yielding a superior SCC-ERGAS trade-off. The visual comparison shows that our network generates clearer edges and fewer color artifacts.

\bibliographystyle{IEEEbib}
\bibliography{strings,refs}

\end{document}